\documentclass[lettersize,journal]{IEEEtran}
\usepackage{amsmath,amsfonts}
\usepackage{amsfonts,amssymb}

\usepackage[linesnumbered,ruled,vlined]{algorithm2e} 
\usepackage{algpseudocode}
\usepackage{algorithm2e}

\SetKwInput{KwInput}{Input}                
\SetKwInput{KwOutput}{Output}              

\usepackage{array}
\usepackage[caption=false,font=normalsize,labelfont=sf,textfont=sf]{subfig}
\usepackage{textcomp}
\usepackage{stfloats}
\usepackage{url}
\usepackage{verbatim}
\usepackage{graphicx}
\usepackage{cite}
\usepackage{multirow}
\usepackage{threeparttable} 

\usepackage{booktabs}
\usepackage{arydshln}
\usepackage[misc]{ifsym}

\newtheorem{definition}{Definition}

\usepackage{color}

\begin{document}

\title{Inductive Subgraph Embedding for Link Prediction}

\author{Chunyu Miao, Chenxuan Xie, Jiajun Zhou, Shanqing Yu, Lina Chen, Qi Xuan,~\IEEEmembership{Senior Member,~IEEE}

\thanks{This work was supported by the Key Project of Regional Innovation and Development Joint Fund of National Natural Science Foundation of China under Grant U22A2025.
\emph{(Co-corresponding authors: Lina Chen, Jiajun Zhou.)}}
\thanks{C. Miao and L. Chen are with the College of Mathematics, Physics and Information Engineering, Zhejiang Normal University, Jinhua Zhejiang 310023, China. C. Miao is also with the Key Laboratory of Peace-building Big Data of Zhejiang Province, Hangzhou 310051, China. E-mail:\{cymiao, chenlina\}@zjnu.cn }
\thanks{C. Xie, J Zhou, S. Yu, and Q. Xuan are with the Institute of Cyberspace Security, Zhejiang University of Technology, Hangzhou 310023, China, with the Binjiang Cyberspace Security Institute of ZJUT, Hangzhou 310023, China. Email:\{221122030330, jjzhou, yushanqin, xuanqi\}@zjut.edu.cn}

}
\markboth{Journal of \LaTeX\ Class Files,~Vol.~14, No.~8, August~2021}%
{Shell \MakeLowercase{\textit{et al.}}: A Sample Article Using IEEEtran.cls for IEEE Journals}


\maketitle

\begin{abstract}
Link prediction, which aims to infer missing edges or predict future edges based on currently observed graph connections, has emerged as a powerful technique for diverse applications such as recommendation, relation completion, etc.
While there is rich literature on link prediction based on node representation learning, direct link embedding is relatively less studied and less understood.
One common practice in previous work characterizes a link by manipulate the embeddings of its incident node pairs, which is not capable of capturing effective link features.
Moreover, common link prediction methods such as random walks and graph auto-encoder usually rely on full-graph training, suffering from poor scalability and high resource consumption on large-scale graphs.
In this paper, we propose Inductive Subgraph Embedding for Link Prediciton (SE4LP) --- an end-to-end scalable representation learning framework for link prediction, which utilizes the strong correlation between central links and their neighborhood subgraphs to characterize links.
We sample the ``link-centric induced subgraphs'' as input, with a subgraph-level contrastive discrimination as pretext task, to learn the intrinsic and structural link features via subgraph classification.
Extensive experiments on five datasets demonstrate that SE4LP has significant superiority in link prediction in terms of performance and scalability, when compared with state-of-the-art methods.
Moreover, further analysis demonstrate that introducing self-supervision in link prediction can significantly reduce the dependence on training data and improve the generalization and scalability of model.
The source code will be available online.
\end{abstract}


\section{Introduction}
Graph Representation Learning (GRL), recently attracting considerable attention, aims to convert discrete graph structures into low-dimensional spaces, preserving essential structural information and properties as continuous vector representations or embeddings. These embeddings, adaptable to specific tasks, enhance various downstream applications; node embeddings facilitate node-level analytics such as node classification~\cite{kipf2017semi} and node clustering~\cite{bo2020structural}, while whole-graph embeddings are instrumental in graph-level tasks like graph classification~\cite{dai2016discriminative}. So far, GRL has spurred advancements in several domains, including social networks~\cite{perozzi2014deepwalk}, biochemical analysis~\cite{subramonian2021motif}, knowledge graphs~\cite{hao2019universal}, etc.

While there is rich studies on node and whole-graph representation learning, GRL for link is relatively less studied and less understood.
One common practice in previous works~\cite{grover2016node2vec,kipf2016variational} characterizes a link by manipulate the embeddings of its incident node pairs, following a ``nodes to link'' (abbreviated as ``node2link'') pattern.
For instance, Node2Vec~\cite{grover2016node2vec} utilizes biased random walks and Word2Vec optimization to learn node representations, and further scores the existence of links via similarity computation.
Graph Auto-Encoder (GAE)~\cite{kipf2016variational,pan2018adversarially} follows the same pattern for link prediction, except that it utilizes the strategy of graph structure reconstruction when learning node representations.
Such ``node2link'' pattern suffers from significant shortcomings. 
Firstly, it fails to accurately capture the intricate interaction information between nodes, predominantly focusing on node properties rather than the distinct characteristics of links, thereby compromising its effectiveness in downstream link prediction tasks. 
Secondly, prevalent GRL methods with this pattern~\cite{perozzi2014deepwalk,grover2016node2vec,kipf2017semi,velivckovic2018graph,velickovic2018deep} typically rely on full-graph learning, i.e., they generally accept entire graphs as input and perform full-graph training for feature extraction, which exhibits poor scalability and substantial resource consumption when handling large-scale graphs, further hindering efficient link representation generation.

Considering \emph{the limited expressiveness of link representations generated via the ``node2link'' pattern} and \emph{the scalability issues plaguing current universal link prediction methods based GRL}, a critical question emerges: \emph{how can a universal and scalable link representation learning framework be developed to facilitate link prediction?}

As we known, links (interactions, relations) inherently indicate correlations between nodes, and leveraging primary interaction patterns between node pairs is crucial for developing expressive link representations. 
When considering two nodes connected by a link in a graph, message propagation between them can occur through both direct (i.e., edge) and indirect interactions (e.g., common neighbors, shortest paths), all encapsulated within a subgraph surrounding this link. 
In this work, we introduce the ``subgraph2link'' pattern following the facts: 
(1) subgraph consisting of target link and their local neighborhood information is informative and plays a critical role to provide structure contexts for link representation learning;
(2) subgraph serves as the receptive field of central link, significantly smaller than the entire graph, thereby presenting a viable strategy to eliminate the necessity for full-graph training. 
Inspired by the existing research in structural link learning~\cite{zhang2018link} and scalable graph learning~\cite{jiao2020sub,hamilton2017inductive}, we propose Inductive \textbf{S}ubgraph \textbf{E}mbedding for \textbf{L}ink \textbf{P}rediction (SE4LP).
In SE4LP framework, we first sample subgraphs around target links, then design a self-supervised subgraph contrast task to better characterize the similarities and differences of interaction patterns around different links, and finally predict the existence of central links via subgraph classification. Our SE4LP jointly trains subgraph contrast and subgraph classification tasks in an end-to-end manner, further achieving high-performance link prediction.
The major contributions of our work are summarized as follows:
\begin{itemize}
  \item \textbf{SE4LP}: We propose an end-to-end scalable link representation learning framework via subgraph contrast, which utilizes informative local subgraphs surrounding links to learn highly expressive link representations.
  \item \textbf{Scalability}: We take the receptive field subgraphs extracted from a batch of links as the input during each training step, so as to learn link representation efficiently and make our SE4LP scale well on large-scale graphs. 
  \item \textbf{Effectiveness}: Extensive experiments demonstrate the superiority of our framework in terms of performance and scalability on link prediction. Furthermore, introducing self-supervised learning to link prediction help to learn effective link representation with less training samples.
\end{itemize}

\section{Preliminaries}
Let $\mathcal{G} =(\mathcal{V} ,\mathcal{E})$ be an undirected and unweighted graph, where $\mathcal{V} =\{v_i \mid i=1,2,\cdots ,N\}$ and $\mathcal{E} \subseteq \mathcal{V} \times  \mathcal{V} $ represent the sets of nodes and edges respectively. 
We regard an edge in graph $(v_i,v_j)\in \mathcal{E}$ as a positive link while those nonexistent edges $(v_i,v_j)\notin \mathcal{E}$ are treated as negative links.
Generally, graph structural data consist of two features: a node attribute matrix $\boldsymbol{X} \in \mathbb{R}^{N\times F}$ and an adjacency matrix of graph topology $\boldsymbol{A} \in \mathbb{R}^{N\times N}$, where $\boldsymbol{x}_i\in\mathbb{R}^F$ is the $F$-dimensional feature vector of node $v_i$, $\boldsymbol{A} _{ij} =1$ if $(v_i,v_j) \in \mathcal{E} $ and 0 otherwise.
We use a diagonal degree matrix $\boldsymbol{D} \in \mathbb{R}^{N\times N}$ to define the degree distribution of $\boldsymbol{A} $, and $\boldsymbol{D}_{ii} = \sum_{j=0}^{N-1} \boldsymbol{A}_{ij}$.

The problem we consider in this work is link prediction via subgraph contrastive representation learning.
Given a graph $\mathcal{G} =(\mathcal{V}, \mathcal{E}, \boldsymbol{X})$, our goal is to learn an encoder $\boldsymbol{h} = f_\theta  (\boldsymbol{X}, \boldsymbol{A})$ which maps a subgraph centered on target link to a vector as the link representation, following the ``subgraph2link'' pattern. 

\begin{figure*}[htp]
	\centering
  \includegraphics[width=\textwidth]{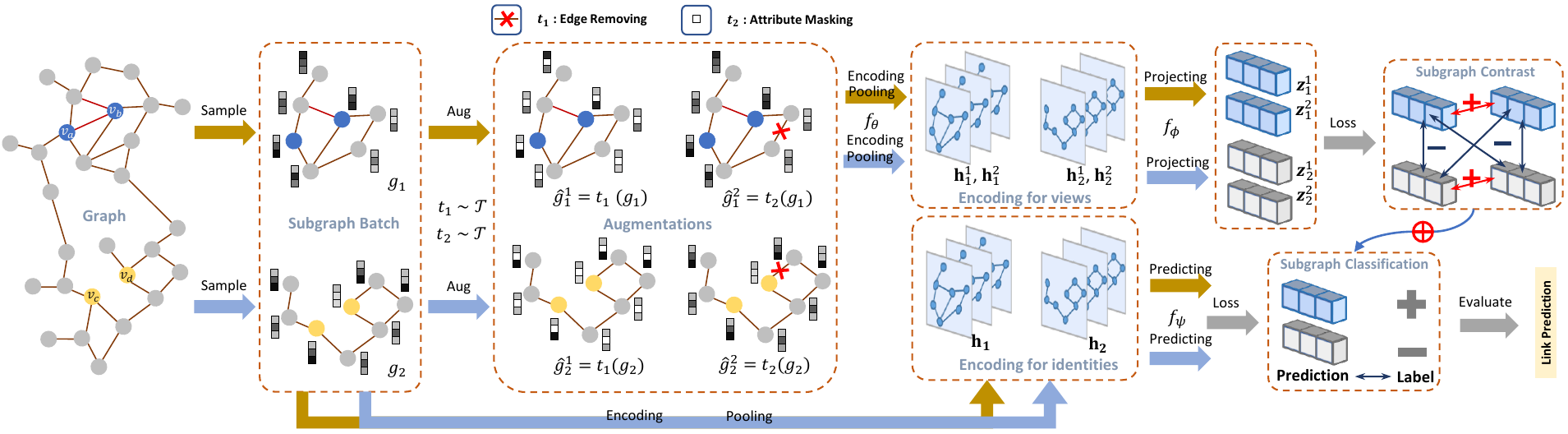}
  \caption{The architecture of SE4LP. The complete workflow proceeds as follows:
           1) extracting and sampling subgraph centered on target links to form input batchs;
           2) applying two augmentation operators on each subgraph to generate two correlated views;
           3) feeding these augmented subgraphs into the GNN encoder and projection head to generate subgraph embeddings as link representations;
           4) maximizing the consistency between two augmented views of subgraphs via a subgraph-level contrast;
           5) mapping subgraph representations to labels reflecting link existence via a subgraph predictor.
           }
  \label{fig:framework}
\end{figure*}

\section{Methodology}
In this section, we give the details of the proposed framework SE4LP, as schematically depicted in Figure~\ref{fig:framework}.
Our framework is mainy composed of the following components:
(1) a subgraph extractor which captures the subgraphs centered on target links from the graph topology;
(2) a subgraph augmentor which generates a series of variant graph views using various transformations on attributes and topology of subgraphs;
(3) a GNN-based encoder which learns graph-level representation for generated graph views;
(4) a subgraph-level contrast maximizes the consistency between two augmented views of the same subgraph;
(5) a subgraph predictor mapping the subgraph representations to labels reflecting link existence.
Next, we describe the details of each component. 


\subsection{Subgraph Extraction and Sampling} 

We name the subgraph centered on link as \emph{link-centric induced subgraph} (abbreviated as ``\emph{lsg}'') and give the definition.
\begin{definition}
  (\emph{Link-centric Induced Subgraph, lsg}).\  For a graph $\mathcal{G} =(\mathcal{V} ,\mathcal{E})$, given target link $l = (v_i, v_j)$ where $v_i, v_j \in \mathcal{V} $, the $h$-hop link-centric induced subgraph for link $l$ is the subgraph ${g}_{l}^h$ induced from $\mathcal{G}$ by the set of nodes $\cup_{v\in (v_i,v_j)} \{v_k \mid d(v_k, v) \leq h\}$.
\end{definition}
Note that ${g}_{l}^h$ of link $l = (v_i, v_j)$ contains all the $h$-hop neighbors of $v_i$ and $v_j$. 
The informative subgraph patterns can effectively reflect the existence of links between center node pairs, which helps to characterize the link structure.

Furthermore, we know that a $h$-layer GNN expands the receptive field by one-hop during each iteration and after $h$ iterations the features of nodes within $h$-hops will be aggregated.
When it comes to a deeper model, the size (i.e., the number of nodes) of receptive field subgraph grows exponentially with layers, which results in neighborhood explosion problem.
In this work, we use subgraph sampling to control the size of \emph{lsg}, even though it can only alleviate this problem to some extent.
Specifically, for a $h$-hop \emph{lsg} of link $l = (v_i,v_j)$, we select the top-$K_1$ important 1-hop neighbors (with node degree value) for $v_i$ and $v_j$ respectively, and again select the top-$K_2$ important 1-hop neighbors for each selected node at hop 1, and recursive ones in the downstream hops.
The recursive sampling can be formulated as follows:
\begin{equation}
  \mathcal{V}_{t} = \bigcup_{v\in \mathcal{V}_{t-1}} \ \mathsf{topK}(\mathcal{N}_v, K_{t}, \boldsymbol{D}[\mathcal{N}_v,\mathcal{N}_v]),
\end{equation}
where $\mathcal{V}_{t}$ is the set of nodes sampled at hop $t$, $\mathcal{N}_v$ is the 1-hop neighborhs set of node $v$, $K_{t}$ is the sampling number at hop $t$, $\boldsymbol{D} [\mathcal{N}_v,\mathcal{N}_v]$ is the degree sequence of nodes in $\mathcal{N}_v$, and $\mathsf{topK}$ is the function that returns the nodes of top-$K$ largest degree values. 
$\mathcal{V}_0$ is initialized as $\{v_i,v_j\}$. 
After $h$ iterations, the set of nodes sampled from the $h$-hop \emph{lsg} is $\mathcal{V}_l = \cup_{t=0}^{h} \mathcal{V}_t$.
The sampled \emph{lsg} (abbreviated ``\emph{slsg}'' in this paper) $g_l = (\boldsymbol{A}_l, \boldsymbol{X}_l)$ is induced by $\mathcal{V}_l$, and its adjacency matrix $\boldsymbol{A}_l$ and feature matrix $\boldsymbol{X}_l$ are denoted respectively as
\begin{equation}
  \boldsymbol{X}_l = \boldsymbol{X}[\mathcal{V}_l, :] \ , \quad  \boldsymbol{A}_l = \boldsymbol{A}[\mathcal{V}_l, \mathcal{V}_l] \ .
\end{equation}
Note that we remove the edge between $v_i$ and $v_j$ if $l=(v_i,v_j)$ is a positive link.
Finally, we anonymize the \emph{slsg} $g_{l}$ by relabeling its nodes to be $\{1,2, \cdots, |\mathcal{V}_l|\}$, in arbitrary order.
For a set of links $L_\textit{emb}$ to be embedded, we get \emph{slsg} for each link in $L_\textit{emb}$ and all \emph{slsg} form a set: $\mathcal{D} = \{g_{i} \mid i=1,2,\cdots, |L_\textit{emb}|\}$.

\subsection{Subgraph Contrastive Learning for Link Representation}
To learn highly expressive link representations, SE4LP utilizes subgraph contrast as a pretext task to jointly train a GNN encoder $f_\theta $.
During subgraph contrast, for each \emph{slsg} $g_{i}$, its two correlated views $\hat{g}_i^1$ and $\hat{g}_i^2$ are generated by undergoing two augmentation operators $t_1$ and $t_2$, where $\hat{g}_i^1 = t_1(g_{i})$ and $\hat{g}_i^2 = t_2(g_{i})$.
The correlated augmented views are fed into a GNN encoder $f_\theta$ with pooling layer, producing the whole subgraph representations $\boldsymbol{h}_i^1$ and $\boldsymbol{h}_i^2$, which are then mapped into a contrast space via a projection head $f_\phi $, yielding $\boldsymbol{z}_i^1$ and $\boldsymbol{z}_i^2$.
Note that $\theta$ and $\phi$ are the parameters of graph encoder and projection head respectively.
The representation of a \emph{slsg}, $\boldsymbol{h}$, is treated as the representation of its central link, following a ``subgraph2link'' pattern.
Finally, the goal of subgraph-level contrast is to maximize the consistency between two correlated augmented views of subgraphs in the contrast space:
\begin{equation}\label{eq:loss-all}
  \mathcal{L}_\textit{self} = \frac{1}{n} \sum_{i=1}^n  \mathcal{L}_i ,
\end{equation}
where $n$ is the number of subgraphs in a batch (i.e., batch size). The loss for each subgraph $\mathcal{L}_i$ can be computed as:
\begin{equation}\label{eq:loss-one}
  \mathcal{L}_i = -\log \frac{e^{\mathrm{s}\left(\boldsymbol{z}_{i}^{1}, \boldsymbol{z}_{i}^{2}\right) / \tau}}{\sum_{j=1, j \neq i}^{n} e^{\mathrm{s}\left(\boldsymbol{z}_{i}^{1}, \boldsymbol{z}_{j}^{2}\right) / \tau}},
\end{equation}
where $\mathrm{s}(\cdot, \cdot)$ is the cosine similarity function having $\mathrm{s}(\boldsymbol{z}_{i}^{1}, \boldsymbol{z}_{i}^{2}) = {\boldsymbol{z}_{i}^{1}}^\top\cdot \boldsymbol{z}_{i}^{2} / \| \boldsymbol{z}_{i}^{1}\| \| \boldsymbol{z}_{i}^{2}\|$, and $\tau$ is the temperature parameter.
The two correlated views $\boldsymbol{z}_i^1$ and $\boldsymbol{z}_i^2$ of \emph{slsg} $g_{i}$ are treated as positive pair while the rest view pairs in the batch are treated as negative pairs.
The objective aims to maximize the consistency of positive pairs as opposed to negative ones.
Note that here we use an asymmetrtic and simplified loss compared to the SimCLR loss~\cite{chen2020simple}, i.e., we generate negative pairs by only treating view 1 ($\boldsymbol{z}_i^1$) as the anchor and contrasting with view 2 ($\boldsymbol{z}_j^2$) of all other subgraphs, as shown in Eq.~(\ref{eq:loss-one}).

\subsection{Graph Augmentation}
Contrastive learning relies heavily on well-designed data augmentation strategies for view generation.
In this paper, we use two existing augmentation methods, \emph{Attribute Masking} and \emph{Edge Removing}~\cite{zhou2022data,zhou2020m,zhu2021graph}, and design two novel augmentation techniques, \emph{Attribute Similarity} and \emph{KNN Graph}.

\subsubsection{Attribute Similarity}
This augmentor builds new node features based on node similarity. 
Specifically, the new node feature matrix is actually the node similarity matrix $\boldsymbol{S}\in \mathbb{R}^{N\times N}$, in which each entry $\boldsymbol{S}_{ij}$ represents the similarity between node $v_i$ and $v_j$, and can be calculated by $S_{ij} = \boldsymbol{x}_i^\top \boldsymbol{x}_j$.
\begin{equation}
  \hat{\boldsymbol{X}} = t_\textit{AM}(\boldsymbol{S} ) = t_\textit{AM}(\boldsymbol{X}\boldsymbol{X}^\top ).
\end{equation}

\subsubsection{KNN Graph}
This augmentor builds new adjacency matrix based on feature similarity.
For each node with feature $\boldsymbol{S}_i$, we find its top $K_T$ similar samples as neighbors and set edges to connect it and its neighbors, formulated as $\hat{\boldsymbol{a} }_i = \mathsf{bT}(\boldsymbol{S}_i)$ where $\mathsf{bT}$ is the function that binarizes elements in a vector by setting the largest $K_T$ elements as 1 and other elements as 0.
The resulting adjacency matrix $\hat{\boldsymbol{A}}$ can be computed as

\begin{equation}
  \hat{\boldsymbol{A}} =t_\textit{KG}(\boldsymbol{S}) 
    =[\mathsf{bT}(\boldsymbol{S}_1); \cdots; \mathsf{bT}(\boldsymbol{S}_N)]^\top.
\end{equation}

\subsection{Model Training}
We achieve link prediction by a subgraph label predictor $f_\psi$, which maps subgraph representations to labels reflecting link existence, yielding a classification loss:
\begin{equation}
  \mathcal{L}_\textit{pred} = -\frac{1}{n}\sum_{i=1}^N  y_i \cdot \log (f_\psi (\boldsymbol{h}_i)), 
\end{equation} 
where $\mathcal{L}_\textit{pred}$ is the cross entropy loss.
The subgraph contrast is treated as pretext task, and the encoder in SE4LP is jointly trained with the pretext and subgraph classification tasks. 
The loss function consists of both the self-supervised and classification task loss functions, as formularized below:
\begin{equation}
  \mathcal{L} =\mathcal{L}_\textit{pred} + \lambda \cdot \mathcal{L}_\textit{self} \ ,
\end{equation}
where $\lambda$ controls the contribution of self-supervision term.

\begin{table*}
  \renewcommand\arraystretch{1.4}
  \large
  \centering
  \caption{Performance on link prediction task reported in Area Under Curve (AUC) and Average Precision (AP) measures.}
  \label{tb:lp-result}
  \resizebox{\textwidth}{!}{%
  \begin{threeparttable}
  \begin{tabular}{clcccccccccc} 
  \hline\hline
  \multicolumn{2}{c}{\multirow{2}{*}{Method}}                & \multicolumn{2}{c}{Cora}                                                                                                  & \multicolumn{2}{c}{Citeseer}                                                                                              & \multicolumn{2}{c}{Pubmed}                                                                                                 & \multicolumn{2}{c}{Facebook}                                                                                              & \multicolumn{2}{c}{Github}                            \\ 
  \cline{3-12}
  \multicolumn{2}{c}{}                                       & AUC (\%)                                                    & AP (\%)                                                     & AUC (\%)                                                    & AP (\%)                                                     & AUC (\%)                                                    & AP (\%)                                                      & AUC (\%)                                                    & AP (\%)                                                     & AUC (\%)                                                    & AP (\%)                      \\ 
  \hline
  \multirow{4}{*}{\rotatebox{90}{{Heuristics}}}   
  &\quad CN                                                  &56.19$\pm$\footnotesize{0.099}                               &63.08$\pm$\footnotesize{0.059}                               &58.76$\pm$\footnotesize{0.095}                               &65.31$\pm$\footnotesize{0.060}                               &65.53$\pm$\footnotesize{0.050}                               &67.54$\pm$\footnotesize{0.034}                                &88.70$\pm$\footnotesize{0.076}                               &91.72$\pm$\footnotesize{0.045}                               &67.87$\pm$\footnotesize{0.105}                               &73.26$\pm$\footnotesize{0.069}\\ 
  &\quad Salton                                              &56.85$\pm$\footnotesize{0.094}                               &61.73$\pm$\footnotesize{0.065}                               &59.32$\pm$\footnotesize{0.090}                               &63.79$\pm$\footnotesize{0.065}                               &64.78$\pm$\footnotesize{0.057}                               &66.12$\pm$\footnotesize{0.047}                                &88.61$\pm$\footnotesize{0.077}                               &91.07$\pm$\footnotesize{0.054}                               &62.84$\pm$\footnotesize{0.085}                               &60.69$\pm$\footnotesize{0.054}\\ 
  &\quad AA                                                  &57.33$\pm$\footnotesize{0.087}                               &59.36$\pm$\footnotesize{0.081}                               &58.97$\pm$\footnotesize{0.089}                               &60.90$\pm$\footnotesize{0.083}                               &65.26$\pm$\footnotesize{0.049}                               &67.49$\pm$\footnotesize{0.035}                                &87.24$\pm$\footnotesize{0.093}                               &91.18$\pm$\footnotesize{0.060}                               &67.69$\pm$\footnotesize{0.108}                               &75.23$\pm$\footnotesize{0.085}\\ 
  &\quad RA                                                  &57.77$\pm$\footnotesize{0.090}                               &64.75$\pm$\footnotesize{0.055}                               &59.11$\pm$\footnotesize{0.091}                               &65.88$\pm$\footnotesize{0.058}                               &65.27$\pm$\footnotesize{0.047}                               &67.78$\pm$\footnotesize{0.032}                                &85.08$\pm$\footnotesize{0.101}                               &90.43$\pm$\footnotesize{0.057}                               &67.56$\pm$\footnotesize{0.078}                               &77.14$\pm$\footnotesize{0.050}\\ 
  \hline                                                                                        
  \multirow{7}{*}{\rotatebox{90}{{Unsupervised}}}                                               
  &\tnote{$\diamond$}\quad DeepWalk                          &88.14$\pm$\footnotesize{0.055}                               &87.87$\pm$\footnotesize{0.045}                               &85.67$\pm$\footnotesize{0.052}                               &86.11$\pm$\footnotesize{0.044}                               &90.88$\pm$\footnotesize{0.021}                               &87.48$\pm$\footnotesize{0.026}                                &87.65$\pm$\footnotesize{0.012}                               &85.41$\pm$\footnotesize{0.011}                               &81.25$\pm$\footnotesize{0.013}                               &80.25$\pm$\footnotesize{0.012}\\
  &\tnote{$\diamond$}\quad Node2Vec                          &88.65$\pm$\footnotesize{0.058}                               &89.62$\pm$\footnotesize{0.049}                               &87.36$\pm$\footnotesize{0.065}                               &88.15$\pm$\footnotesize{0.059}                               &90.60$\pm$\footnotesize{0.021}                               &89.29$\pm$\footnotesize{0.027}                                &85.64$\pm$\footnotesize{0.022}                               &85.25$\pm$\footnotesize{0.028}                               &80.49$\pm$\footnotesize{0.019}                               &79.62$\pm$\footnotesize{0.016}\\                                                    
  \cdashline{2-12}
  &\tnote{$\diamond$}\quad GAE                               &93.79$\pm$\footnotesize{0.038}                               &93.43$\pm$\footnotesize{0.038}                               &92.63$\pm$\footnotesize{0.013}                               &93.50$\pm$\footnotesize{0.016}                               &91.93$\pm$\footnotesize{0.051}                               &91.84$\pm$\footnotesize{0.051}                                &OOM                                                          &OOM                                                          &OOM                                                          &OOM                                          \\
  &\tnote{$\diamond$}\quad VGAE                              &94.30$\pm$\footnotesize{0.006}                               &94.60$\pm$\footnotesize{0.082}                               &93.78$\pm$\footnotesize{0.046}                               &94.55$\pm$\footnotesize{0.051}                               &89.36$\pm$\footnotesize{0.056}                               &89.37$\pm$\footnotesize{0.056}                                &OOM                                                          &OOM                                                          &OOM                                                          &OOM                                          \\         
  &\tnote{$\diamond$}\quad ARGA                              &90.27$\pm$\footnotesize{0.067}                               &90.01$\pm$\footnotesize{0.069}                               &89.00$\pm$\footnotesize{0.040}                               &89.60$\pm$\footnotesize{0.039}                               &88.00$\pm$\footnotesize{0.049}                               &88.33$\pm$\footnotesize{0.045}                                &OOM                                                          &OOM                                                          &OOM                                                          &OOM                                          \\
  &\tnote{$\diamond$}\quad ARGVA                             &93.26$\pm$\footnotesize{0.041}                               &93.61$\pm$\footnotesize{0.045}                               &94.03$\pm$\footnotesize{0.013}                               &94.30$\pm$\footnotesize{0.014}                               &90.48$\pm$\footnotesize{0.036}                               &90.32$\pm$\footnotesize{0.035}                                &OOM                                                          &OOM                                                          &OOM                                                          &OOM                                          \\         
  \cdashline{2-12}
  &\tnote{$\diamond$}\quad DGI                               &93.15$\pm$\footnotesize{0.023}                               &92.70$\pm$\footnotesize{0.030}                               &91.84$\pm$\footnotesize{0.014}                               &92.31$\pm$\footnotesize{0.014}                               &91.45$\pm$\footnotesize{0.004}                               &90.87$\pm$\footnotesize{0.005}                                &OOM                                                          &OOM                                                          &OOM                                                          &OOM                                          \\         
  \hline
  \multirow{4}{*}{\rotatebox{90}{{Supervised}}}                                                
  &\tnote{$\star$}\quad SEAL($h=1$)                          &95.46$\pm$\footnotesize{0.007}                               &95.84$\pm$\footnotesize{0.010}                               &91.20$\pm$\footnotesize{0.010}                               &93.11$\pm$\footnotesize{0.008}                               &94.24$\pm$\footnotesize{0.015}                               &92.56$\pm$\footnotesize{0.021}                                &97.83$\pm$\footnotesize{0.018}                           &97.29$\pm$\footnotesize{0.023}                                   &96.27$\pm$\footnotesize{0.019}                               &96.01$\pm$\footnotesize{0.017}\\
  &\tnote{$\star$}\quad SEAL($h=2$)                          &\underline{95.88$\pm$\footnotesize{0.007}}                   &\underline{96.14$\pm$\footnotesize{0.008}}                   &91.35$\pm$\footnotesize{0.012}                               &93.10$\pm$\footnotesize{0.008}                               &96.82$\pm$\footnotesize{0.009}                               &98.11$\pm$\footnotesize{0.011}                                &\textbf{98.58}$\pm$\textbf{\footnotesize{0.001}}         &\textbf{98.71}$\pm$\textbf{\footnotesize{0.001}}                 &\textbf{97.11}$\pm$\textbf{\footnotesize{0.014}}             & \textbf{97.02}$\pm$\textbf{\footnotesize{0.011}}\\
  \cdashline{2-12}                                           

  &\tnote{$\star$}\quad SE4LP($h=1$)                          &\textbf{96.05}$\pm$\textbf{\footnotesize{0.007}}             &\textbf{96.28}$\pm$\textbf{\footnotesize{0.010}}             &\textbf{94.74}$\pm$\textbf{\footnotesize{0.007}}             &\textbf{95.24}$\pm$\textbf{\footnotesize{0.008}}             &\textbf{98.36}$\pm$\textbf{\footnotesize{0.001}}             &\textbf{98.25}$\pm$\textbf{\footnotesize{0.002}}              &\underline{97.94$\pm$\footnotesize{0.002}}               &\underline{97.64$\pm$\footnotesize{0.001}}                       &\underline{96.43$\pm$\footnotesize{0.005}}                   &\underline{96.11$\pm$\footnotesize{0.005}}\\
  &\tnote{$\star$}\quad SE4LP($h=2$)                          &94.33$\pm$\footnotesize{0.008}                               &94.03$\pm$\footnotesize{0.012}                               &\underline{93.38$\pm$\footnotesize{0.006} }                  &\underline{93.67$\pm$\footnotesize{0.008} }                  &\underline{98.35$\pm$\footnotesize{0.004} }                  &\underline{98.18$\pm$\footnotesize{0.005}}                    &97.53$\pm$\footnotesize{0.001}                           &97.24$\pm$\footnotesize{0.001}                                   &96.12$\pm$\footnotesize{0.010}                               &95.86$\pm$\footnotesize{0.015}\\ 
  \hline\hline
  \end{tabular}
  \begin{tablenotes}
    \footnotesize
    \item[$\diamond$] This method generates link representation with ``node2link'' pattern.
    \item[$\star$] This method generates link representation with ``subgraph2link'' pattern.
  \end{tablenotes}
  \end{threeparttable}}
\end{table*}

\section{Experiments}
\subsection{Experimental Setting}
\noindent\textbf{Dataset:} To assess how well our SE4LP can learn highly expressive link representation while keeping high scalability on large-scale graphs, we evaluate SE4LP on publicly available real-world datasets as follows: Cora, Citeseer, Pubmed, Facebook and Github.
For the two large datasets (Facebook and Github), their initial features are not aligned. 
We create tagged documents from feature hash and further process them into 128-dimensional initial features by Doc2Vec algorithm.
Both of the Facebook and Github datasets are available online~\footnote{http://snap.stanford.edu/data/}.

\noindent\textbf{Data Preparation:}
For methods with ``subgraph2link'' pattern, SEAL and SE4LP, we sample the same number of positive and negative links from each dataset, with the proportion of \{40\%, 40\%, 10\%, 10\%, 20\%\} for \{Cora, Citeseer, Pubmed, Facebook, Github\}. 
All links are split into training, validation and testing sets with a proportation of 8:1:1, and we further extract \emph{slsg} for each link.
For heuristics (CN, Salton, AA, and RA), random walks (DeepWalk and Node2Vec), GAEs (GAE, VGAE, ARGA and ARGVA) and contrastive method (DGI) which rely on full-graph training, we randomly sample a certain number of positive links as well as the same number of additionally negative links as testing data, and the remaining partially observed graph is used for training.
We repeat 10-fold cross validation for 5 times and report the average Area Under Curve (AUC) and Average Precision (AP) measures as well as their standard deviations.

\noindent\textbf{Parameter Configuration:}
For SE4LP, we choose the number of hops from $\{1,2\}$ and set both probabilities ($p_A$, $p_x$) associated with data augmentation to 20\%.
We set the default GNN encoder of SE4LP to graph isomorphism network (GIN)~\cite{xu2018powerful}, which is a 3-layers graph convolutional network with hidden dimension of 128, jumping connection, and PReLU activation function.
We return subgraph representations by max pooling.
The batch size $n$, learning rate, temperature parameter $\tau$ and trade-off coefficient $\lambda$ are set to 512, 0.01, 0.2, 0.1, respectively.
We use early stopping with a patience of 20.

For random walks, we set the length of walks to 30, the number of walks to 200, and the context size to 10. 
For Node2Vec, we set the return parameter $p$ and in-out parameter $q$ to 4 and 1, respectively.
For all GAEs, we set the encoder as 2-layer GCN network.
For GAE and VGAE, we construct encoders with a 128-dimensional hidden layer and a 128-dimensional embedding layer for all the experiments.
For ARGA and ARGVA, we construct encoders with a 256-dimensional hidden layer and a 128-dimensional embedding layer for all the experiments and all the discriminators are built with 2 hidden layers(16 and 64 dimensions).
The learning rates are set to \{0.01, 0.01, 0.005, 0.005\} for GAE, VGAE, ARGA and ARGVA, respectively.
For DGI, we construct encoder with a 512-dimensional hidden layer and a 512-dimensional embedding layer for all the experiments.
For other parameters, we use the default setting following~\cite{velickovic2018deep}.
For SEAL, we choose the number of hops from $\{1,2\}$.
For other parameters, we use the default setting following~\cite{zhang2018link}.

\subsection{Evaluation on Link Prediction}
Table~\ref{tb:lp-result} reports the results on link prediction, from which we can see that SE4LP achieves state-of-the-art results with respect to baselines.
Specifically, our SE4LP significantly outperforms heuristic and random walks baselines across all datasets, indicating that the learned subgraph patterns are better at capturing the link properties than manual features or shallow topology features.
When compared to GAEs, our SE4LP surpasses strong baseline: on three citation benchmarks we observe 1.9\%, 0.8\% and 7.0\% relative improvement over best GAE in terms of AUC, respectively.
We also compare with the state-of-the-art supervised model, SEAL, in a variety of hyper-parameter settings.
The results shown in Table~\ref{tb:lp-result} suggest that our model outperforms SEAL (with same hyper-parameter settings) on three citation benchmarks in most cases, and achieves competitive results in Facebook and Github datasets.
Specifically, SC4LP performs better with 1-hop \emph{slsg}, while SEAL achieves better performance with 2-hop \emph{slsg}.
SC4LP beats SEAL when accepts subgraphs with a smaller size ($h=1$)
When compared to SEAL that accepts larger subgraphs ($h=2$), SC4LP still beats SEAL in 2 out of 5 datasets.
These results suggest that SEAL relies on enclosing subgraphs of larger size to learn high-order link features while our SE4LP prefers to find link patterns from small subgraphs.
Actually, SEAL uses Node Labeling trick to highlight the structural information of subgraphs, and our SE4LP outperforms SEAL on benchmarks without \emph{Node Labeling}, indicating the effectiveness of combining ``subgraph2link'' pattern with self-supervised learning for link prediction.

\noindent\textbf{subgraph2link vs. node2link:}
When comparing the two categories of methods with different patterns, the results suggest that methods with ``subgraph2link'' pattern generally perform on par with or better than strong baselines with ``node2link'' pattern.
Moreover, ``subgraph2link'' models are trained on \emph{slsg} while ``node2link'' models are trained on partially observed graph where only the testing links are masked.
In summary, compared with ``node2link'' methods, ``subgraph2link'' methods use relatively less training data and achieve superior performance and generalizability, validating the effectiveness of our proposal, i.e., \emph{local subgraph can provide informative structure contexts for link representation learning}.

\begin{figure}[htp]
	\centering
  \includegraphics[width=\linewidth]{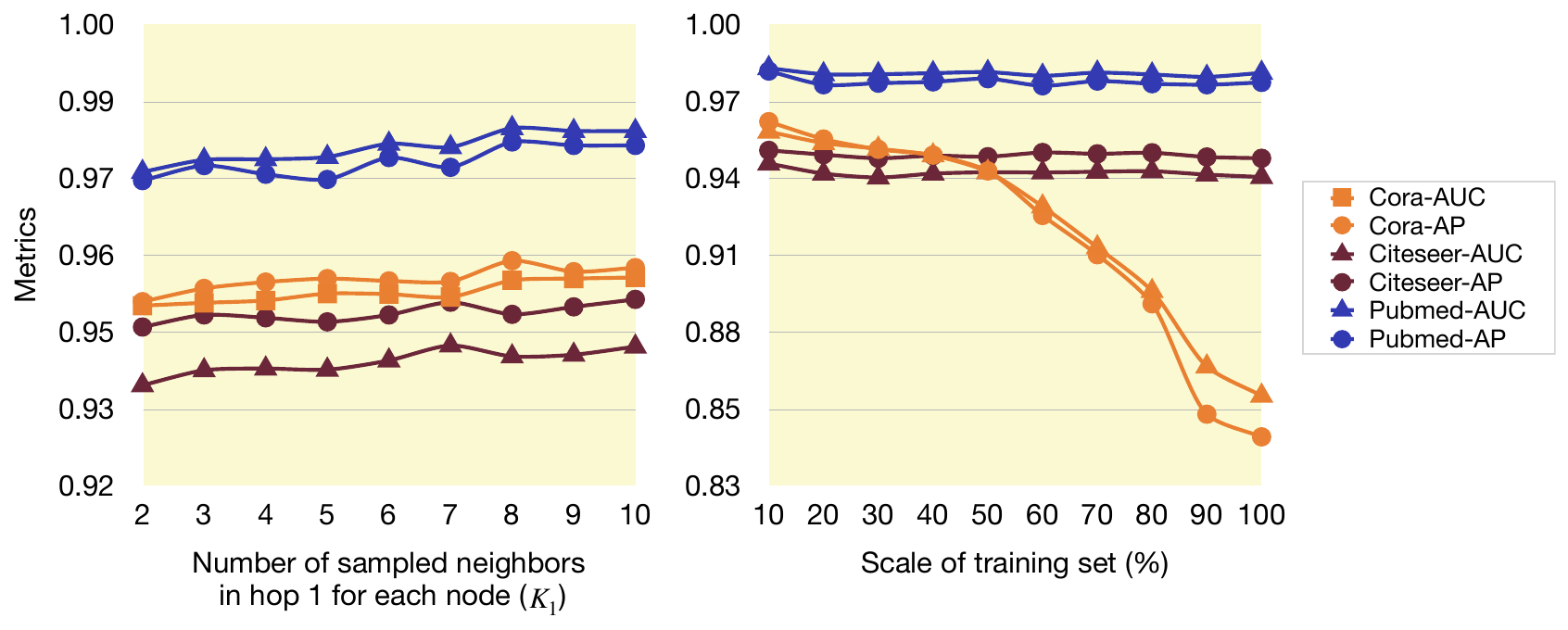}
  \caption{Impact of subgraph size and training set scale.}
  \label{fig: para}
\end{figure}

\subsection{More Analysis}
\noindent\textbf{Impact of Subgraph Size:}
We further investigate the impact of subgraph size in our SE4LP on citation benchmarks.
We first fix $h=1$ and adjust $K_1$ from 2 to 10, and evaluate the results shown in Figure~\ref{fig: para}~(a).
We observe that the performance of SE4LP increases slightly as the size of the subgraph increases, indicating that SE4LP has certain robustness to the variation in subgraph scale.
Moreover, our method can be adapted to subgraphs of different sizes and capture key pattern features which reflect link existence.
As a result, the subgraph sampling module can improve the scalability of SE4LP in large-scale graphs by accepting small subgraphs to achieve highly powerful link prediction.

\noindent\textbf{Impact of Self-supervised Learning in Link Prediciton:}
We further investigate the impact of contrastive self-supervision in our framework, which can be measured by the scale of training data.
Specifically, we cut the training set from 10\% to 90\%, and observe the performance of SE4LP on link prediction, as shown in Figure~\ref{fig: para}~(b).
As we can see, SE4LP always keeps stable performance with the reduction of training data on Citeseer and Pubmed datasets.
Moreover, in our experiments, SE4LP uses 10\% $\sim$ 40\% positive links for training, while GAEs and random walks use ``partially observed graph" (80\% positive links) for training. 
As a result, SE4LP can learn effective link representations with less training data, and achieve competitive or even better performance in link prediction.
Generally, link prediction tasks have no shortage of training data because the number of edges in the graph is generally large. 
So we conclude that introducing self-supervision in link representation learning can significantly reduce the dependence on training data and improve the generalization and scalability.

\section{Conclusion}
Graph representation learning for link is relatively less studied and less understood.
In this paper, we study the ``subgraph2link'' patern and propose an end-to-end joint learning framework for learning link representations by contrasting encodings from different view of subgraphs centered on links.
Experiments conducted on five datasets demonstrate that our framework can achieve new SOTA performance in link prediction, and have satisfactory scalability on large-scale datasets.

\bibliographystyle{IEEEtran}
\bibliography{main,IEEEabrv}

\end{document}